\newif\ifshowfig
\def\ifshowfig{\iftrue}

\newif\ifsubmit
\def\ifsubmit{\iftrue}

\documentclass[epj,a4paper]{svjour}

\usepackage{graphicx}
\usepackage{bm,bbm}
\usepackage{amssymb,amsfonts,amsmath}
\usepackage[english]{babel}
\usepackage[utf8]{inputenc}
\usepackage{math4phy}
\usepackage{xcolor}
\usepackage{hyperref}
\usepackage{ulem}
\usepackage{braket}
\usepackage{cite}

\begin{document}

%%% Headlines %%%%%%%%%%%%%%%%%%%%%%%%%%%%%%%%%%%%%%%%%%%%%%%%%%%%%%%%%%%%%%
\title{Dynamics of Solitons in the One-Dimensional Nonlinear
  Schr\"odinger Equation}
\author{Tobias Ilg \and Ramona Tsch\"uter \and Andrej Junginger \and
  J\"org Main \and G\"unter Wunner}
\institute{Institut f\"ur Theoretische Physik 1, Universit\"at Stuttgart, 
  70550 Stuttgart, Germany}
\date{\today}
\authorrunning{T.~Ilg \etal}

%%% user-defined commands %%%%%%%%%%%%%%%%%%%%%%%%%%%%%%%%%%%%%%%%%%%%%%%%%%
\newcommand{\ie}{i.\,e.}
\newcommand{\eg}{e.\,g.}
\newcommand{\cf}{cf.}
\newcommand{\etal}{\textsl{et~al.}}
\newcommand{\FIG}{Fig.}
\newcommand{\FIGS}{Figs.}
\newcommand{\SEC}{Sec.} \newcommand{\SECS}{Secs.}
\newcommand{\EQ}{Eq.}
\newcommand{\EQS}{Eqs.}
\newcommand{\REF}{Ref.}
\newcommand{\REFS}{Refs.}
\newcommand{\comment}[1]{\textcolor{orange}{\textsf{[#1]}}}   %%%%%%%%%%%%
\definecolor{mygreen}{RGB}{0,200,0} 
\newcommand{\Ng}{{N_\text{g}}}
\newcommand{\Ns}{{N_\text{s}}}
\newcommand{\transpose}{\mathsf{T}}
\newcommand{\zz}{\vec{z}}
\newcommand{\prest}{\bar{p}}
\newcommand{\ue}{\mathrm{e}}
\newcommand{\ud}{\mathrm{d}}
\definecolor{myred}{RGB}{200,0,0}
\definecolor{mygreen}{RGB}{0,150,0}

\ifsubmit
  \newcommand{\STRIKE}[1]{}
  \newcommand{\EDIT}[1]{#1}
  \newcommand{\strike}[1]{}
  \newcommand{\strikeeq}[1]{}
  \newcommand{\edit}[1]{#1}
\else
  \newcommand{\strike}[1]{\textcolor{red}{\sout{[#1]}}}
  \newcommand{\strikeeq}[1]{#1}
  \newcommand{\edit}[1]{\textcolor{mygreen}{#1}}
  \newcommand{\STRIKE}[1]{{\color{myred}\sout{#1}}}
  \newcommand{\EDIT}[1]{{\color{mygreen}#1}}
\fi
%%%%%%%%%%%%%%%%%%%%%%%%%%%%%%%%%%%%%%%%%%%%%%%%%%%%%%%%%%%%%%%%%%%%%%%%%%%%

%%% Abstract %%%%%%%%%%%%%%%%%%%%%%%%%%%%%%%%%%%%%%%%%%%%%%%%%%%%%%%%%%%%%%%
\abstract{%
We investigate bright solitons in the one-dimensional Schrödinger 
equation in the framework of an extended variational approach. We apply the 
latter to the stationary ground state of the system as well as to coherent 
collisions between two or more solitons. 
Using coupled Gaussian trial wave functions, we demonstrate that the 
variational 
approach is  a powerful method to calculate the soliton dynamics. 
This method has the advantage that it is computationally faster compared to 
numerically exact grid calculations. 
In addition, it goes far beyond the capability of analytical ground state 
solutions, because the variational approach provides the ability to treat 
excited 
solitons as well as dynamical interactions between different wave packets.
To demonstrate the power of the variational approach, we calculate the 
stationary ground state of the soliton and compare it with the analytical 
solution 
showing the convergence to the exact solution. Furthermore, we extend our 
calculations to nonstationary solitons by investigating \strike{the breathing 
oscillations of excited solitons and the} coherent collisions of several wave 
packets \edit{in both the low- and high-energy regime}.
Comparisons of the variational approach with numerically exact simulations on 
grids reveal excellent agreement in the high-energy regime \edit{while 
deviations can be observed for low energies}.
}

%%% Pacs %%%%%%%%%%%%%%%%%%%%%%%%%%%%%%%%%%%%%%%%%%%%%%%%%%%%%%%%%%%%%%%%%%%

%%% Title %%%%%%%%%%%%%%%%%%%%%%%%%%%%%%%%%%%%%%%%%%%%%%%%%%%%%%%%%%%%%%%%%%
\maketitle

% \STRIKE{Entfernter Text}
% \EDIT{Neuer Text}

%%% Introduction %%%%%%%%%%%%%%%%%%%%%%%%%%%%%%%%%%%%%%%%%%%%%%%%%%%%%%%%%%%
\section{Introduction}

Solitons are nondispersive wave packets and they are commonly known to exist as 
a general phenomenon in nonlinear wave dynamics in various fields 
\cite{Eichler1,Eichler2,Tikhonenkov2008,Khaykovich2002,Strecker2002, 
Paredes2016, Anderson1985, Syafwan2012}. 
The fact that the wave packet keeps its shape arises from two counter-acting 
influences.
A broadening effect is compensated by an attractive interaction, so that there 
is an intermediate state where the wave packet is stable.
Beyond the existence of soliton wave packets in various systems their  
nonstationary dynamics is of high interest. 
This can \eg\ be collective oscillations of slightly excited solitons or 
collisions between them.

In this paper, we focus on quantum mechanical solitons in Bose-Einstein 
condensates (BECs).
Here, Heisenberg's uncertainty principle, \ie\ the kinetic energy term of the 
condensate, favors a broadening of the wave packet. 
By contrast, an attractive interaction, \eg\ attractive s-wave scattering 
or an additional long-range interaction \cite{Wunner2008}, favors a 
shrinking 
condensate wave function.
As a consequence, the BEC soliton exists when these terms compensate each other.
In general, one distinguishes bright and dark solitons, where bright solitons 
exhibit an increased density resulting from attractive interactions while dark 
solitons are density minima due to repulsive interactions \cite{Pethick2008}.

Mathematically, the condensate wave function obeys the time-dependent 
Gross-Pitaevskii equation (GPE) 
\cite{Gross61a,Pitaevskii61a,Dalfovo1999,Ueda2010}, a 
nonlinear extension of the ordinary Schrödinger equation in which the nonlinear 
interaction term results from a mean-field approximation of the BEC. 
In general, several methods can be used to solve the GPE in order to determine 
the condensate's ground state and its dynamics.
These are \eg~the direct numerical integration in the one-dimensional case, the 
discretization of the wave function on a grid, as well as 
the description of the condensate within a variational approach 
\cite{Eichler1,Eichler2},

From the theoretical point of view, the one-dimensional BEC-soliton is of 
particular interest, because in this case the GPE can be solved analytically 
and the wave function of the condensate can be expressed by a simple analytic 
function \cite{Pethick2008}.
This wave function, however, can only describe the ground state of the soliton, 
while dynamical solutions beyond a simple translation in space, excitations, or 
collisions are not accessible within this analytical approach.
Thus, the latter situations require a different method and in this paper we 
describe a promising possibility using an extended variational approach based 
on coupled Gaussian wave functions.
Such variational approaches have already been applied successfully to different 
questions arising in the field of BECs ranging from the simple reproduction of 
its ground state wave function 
\cite{Huepe2003, PerezGarcia1996, PerezGarcia1997, Yi2000, Yi2001, Parker2009, 
Muruganandam2012}, 
over issues of stability
\cite{Kreibich2012,Kreibich2013a}, ground and excited states of BECs with 
long-range 
$1/r$- or dipolar interactions \cite{Rau2010a,Rau2010b,Rau2010c}, 
to the collapse dynamics of the condensate 
\cite{Huepe1999, Junginger2010, Junginger2012b, junginger2012d, Junginger2013b}.
\edit{
However, in these situations a direct comparison between variational and 
directly numerical approaches has not been performed or has not been possible. 
Therefore, it is one of the goals of this paper to provide direct comparisons of 
these approaches.}

In experiments, the BEC is usually confined in a harmonic trap whose ground 
state wave function is of pure Gaussian form.
The choice of using Gaussian variational approaches is therefore 
straightforward in these cases.
Regarding solitons in the absence of external traps, the situation is  
different, because the condensate wave function exhibits a broad heavy-tail 
distribution.
As a consequence, it is a nontrivial question to what extent a Gaussian-based 
approach is capable of recovering the correct results.
As we will demonstrate below, only few coupled Gaussians are necessary 
to obtain converged wave functions.
In addition to the description of the condensate's ground state, a particular 
advantage of the variational approach lies in its capability to derive equations 
of motion of the condensate dynamics analytically. 
This is due to the fact that its reduction of the degrees of freedom in general 
makes the computational effort significantly smaller -- often by several orders 
of magnitude in the computation time as compared to grid calculations.
Nevertheless, the accuracy of the variational approach is high enough to 
correctly describe the condensate dynamics.
We demonstrate this for the collision of solitons, for which we directly 
compare the collision dynamics of the variational approach with that obtained 
from numerical grid calculations.

Our paper is organized as follows:
In \SEC~\ref{sec:theory}, we present the theoretical description of the 
solitons, review the analytical solution of the ground state, and introduce the
variational approach.
In \SEC~\ref{sec:results}, we demonstrate by comparison with numerically exact 
simulations, the capability of the variational approach to reproduce both the 
stationary as well as dynamical behavior of single and colliding solitons.

%%% Section: Theory %%%%%%%%%%%%%%%%%%%%%%%%%%%%%%%%%%%%%%%%%%%%%%%%%%%%%%%%
\section{Theory} 
\label{sec:theory}

In mean-field approximation, the dynamics of one-dimen\-sional BEC solitons 
with wave function $\psi$ is described by the 
time-depen\-dent GPE
\begin{equation}
 \textup{i}\hbar \frac{\mathrm{d}}{\mathrm{d}t} \psi(x,t) = \left[ - 
\frac{\hbar^2}{2m} \frac{\mathrm{d}^2}{\mathrm{d}x^2} + g |\psi(x,t)|^2 \right] 
\psi(x,t) \,.
\label{eq:GPE}
\end{equation}
Here, $m$ is the mass of the bosons and $g$ describes the scattering 
interaction between the single bosons within an s-wave approximation.
Negative values of $g$ mean an attractive interaction while positive values 
describe a repulsive one.
In this paper, we only consider the case $g<0$.
Due to the absence of any external potential in \EQ~\eqref{eq:GPE}, the 
particles are free in the $x$-direction and the total dynamics is determined 
by the counterplay between Heisenberg's uncertainty principle (\ie~the kinetic 
energy term) and the attraction between the particles.

The free parameter $g<0$ in \EQ~\eqref{eq:GPE} suggests a different 
behavior of the underlying system in dependence of its actual value.
However, using an appropriate system of units, this parameter can be eliminated 
from \EQ~\eqref{eq:GPE}, showing that the underlying physics is the same for 
all $g<0$.
To do so, we measure the action in units of $\hbar$ and the unit 
of mass is set to $m_0=2m$. 
Furthermore, the unit of energy is 
$ E_0={\hbar^2}/({m_0x_0^2})$
with $x_0$ being the unit of length, and the coupling constant is measured in 
terms of
$ g_0={\hbar^2}/({m_0x_0})$.
Since the unit of length, $x_0$, is a free parameter, its value and, thus, 
the value of $g_0$ can always be chosen such that one obtains $g=-1$ in the 
scaled units.
\edit{Without loss of generality the GPE} \strike{The GPE then} simplifies to
\begin{equation}
\textup{i}\frac{\mathrm{d}}{\mathrm{d}t}\psi(x,t)= 
% \underbrace{
\strike{\left[ 
-\frac{\mathrm{d}^2}{\mathrm{d}x^2} - |\psi(x,t)|^2 \right]}
% }_{\equiv H} 
\edit{\hat{H}}\psi(x,t)\strike{\,, }
\label{eq:GPEdimensionslos}
\end{equation} 
\edit{with the Hamiltonian }
\begin{equation}
\edit{\hat{H}=-\frac{\mathrm{d}^2}{\mathrm{d}x^2} - |\psi(x,t)|^2\,.}
\end{equation}
\strike{without loss of generality.}

\subsection{Analytical solution of the soliton's ground state}

As mentioned in the introduction, the stationary case of the 
GPE~\eqref{eq:GPEdimensionslos} can be 
solved analytically.
As one can find \eg~in the book of Pethick and Smith \cite{Pethick2008}, the 
ground state of the soliton is given by
\begin{equation}
\psi(x,t) = \frac{b}{\cosh(ax)} \, \ue^{- \textup{i}\mu t} \,,
\label{eq:PSI1}
\end{equation}
where $\mu<0$ is the chemical potential and the parameters $a$ and $b$ 
determine the width of the soliton and its amplitude through the relations
\begin{equation}
 a^2 = - \mu  \quad \text{and} \quad
 b^2 = -2 \mu 
 \label{eq:ab3}
\end{equation}
with $\mu = -{1}/{16}$ for normalized wave functions.
The mean-field energy of the soliton \eqref{eq:PSI1} is given by the 
expectation 
value
\begin{equation}
E=\left\langle -\frac{\mathrm{d}^2}{\mathrm{d}x^2}-
\frac{1}{2}|\psi(x,t)|^2\right\rangle=-\frac{1}{48} \,,
\label{eq:E}
\end{equation}
where the additional factor $1/2$ is necessary in order to avoid a 
double-counting of the two-particle interactions.
The negative value of the energy shows that the soliton is a bound state.

\subsection{Lattice calculation}

A straightforward way to obtain the time evolution in the excited case
or if several solitons collide
is to 
iterate the wave function on a spacial lattice. 
Therefore, we label the single lattice sites with the variable $j\in\mathbb{Z}$ 
and the distance between them with $\Delta x$.
In order to obtain a time-dependent wave function of an excited soliton one 
then applies the time-evolution operator $U(t,t_0)$ which reads
\begin{equation}
U(t_0+\delta t, t_0)=1-\textup{i} \hat H \delta t + O(\delta t^2)
\end{equation}
for small time steps $\delta t$. 
\edit{
Note that the first-order approximation of the time-evolution operator is not 
norm-con\-ser\-ving in general in contrast to, \eg, a Crank-Nicolson scheme 
\cite{numericalrecipes}. 
However, using sufficiently small time steps $\delta t$, a possible error can be 
kept negligibly small which we verify in addition by a direct calculation of 
the time-dependence of the norm in our simulations.
}
The value $\psi(x_j,t_0+\delta t)$ 
of the wave function at the lattice site $j$ 
is then given by
\begin{align}
\begin{split}
\psi(x_j, t_0+\delta t) &\approx \psi(x_j,t_0) \\
 &+\textup{i}
\left[ 
\frac{\mathrm{d}^2}{\mathrm{d}x^2} + |\psi(x_j, t_0)|^2 
\right]\psi(x_j, t_0)\delta t\,
\end{split}
\label{eq:iteration}
\end{align}
and the second derivative therein can be evaluated via 
\begin{equation}
\frac{\mathrm{d}^2}{\mathrm{d}x^2}\psi(x_j, t_0)=\frac{\psi(x_{j-1}, 
t_0)+\psi(x_{j+1},t_0)-2\psi(x_j,t_0)}{\Delta x^2}\,.
\end{equation}  
With \EQ~\eqref{eq:iteration} one can calculate the wave function at any time, 
and we will use this method to compare the variational approach with 
numerically exact results.

\subsection{Variational approach to the soliton}

The purpose of this paper is to describe the one-dim\-en\-sion\-al soliton 
dynamics within a variational framework. 
In this approach, we parametrize the wave function 
\begin{equation}
\psi(x,t) = \psi(x, \zz(t))
\label{eq:var-approach-general}
\end{equation}
by a set of complex and time-dependent variational parameters $\zz(t)$.
In order to determine the dynamics of the system, we use the McLachlan 
variational principle \cite{Frenkel1934,McLachlan1964}
\begin{equation}
\STRIKE{
\|\textup{i} \dot \psi - \hat H \psi \|^2 \stackrel{\mathrm{!}}= 
\textup{min}\,.
}
\EDIT{
\|\textup{i} \dot \psi - \hat H \psi \|^2 = 
\textup{min}\,.
}
 \label{eq:SGL-Norm}
\end{equation}
\edit{
If the wave function $\psi$ is an exact solution of the 
GPE~\eqref{eq:GPE}, the norm in \EQ~\eqref{eq:SGL-Norm} vanishes and the 
minimum is zero, while it is nonzero for an approximate solution. 
The optimal time-evolution of the trial wave function is obtained by requiring a 
minimum norm
% This is achieved by a varying \EQ~\eqref{eq:SGL-Norm} with respect to 
% $\dot\psi$ 
meaning that the time-evolution of the variational wave function is 
solved with the least possible error. 
The application of the time-depen\-dent variational principle 
\eqref{eq:SGL-Norm} 
to Gaussian functions is well-established in the literature, 
% \cite{Rau2010a, Rau2010b, Rau2010c}, 
so that we only briefly sketch the basic steps in the following.
For a detailed description, we refer the reader to 
\REFS~\cite{Rau2010a, Rau2010b, Rau2010c} and references therein.
}

\edit{
Calculating the minimum of \EQ~\eqref{eq:SGL-Norm} using the general approach 
\eqref{eq:var-approach-general} yields a set of first-order coupled 
differential equations 
}
\begin{equation}
\Braket{
\frac{\partial\psi}{\partial\zz}|
\textup{i}\dot{\psi}+
\frac{\ud^2}{ \ud x^2}\psi 
+ \left|\psi\right|^2 \psi}=0
  \label{eq:TDVP-eq-of-motion}
\end{equation}
which determine the time-evolution of the system. 
In addition, the energy of the system is given by the energy functional 
\begin{equation}
E = \Braket{
-\frac{\ud^2}{ \ud x^2} - 
\frac{1}{2}\left|\psi\right|^2 } \,.
  \label{eq:TDVP-E}
\end{equation}
For our investigations, we use as a trial wave function a superposition of 
Gaussians according to
\begin{align}
\psi = \sum\limits_{n=1}^{\Ng} g_n
\quad \text{with} \quad
g_n = \ue^{-\alpha_{n}x^2+\beta_{n}x+\gamma_ {n}}
\,,
\label{eq:ansatz}
\end{align}
where $\Ng$ is the number of Gaussians used to approximate the soliton(s).
The parameter $\alpha_{n}$ determines the width of each Gaussian, $\gamma_{n}$ 
its weight and phase, and $\beta_{n}$ the velocity and position to the wave 
package. 
Altogether, the total set of variational parameters can be summarized by
$\zz=[\alpha_{1}$,$\beta_{1}$,$\gamma_{1},\ldots$,$\alpha_{\Ng}$,
$\beta_{\Ng}$,$\gamma_{\Ng}]^\transpose$
and we note that by initializing the single Gaussians $g_n$ with appropriate 
different linear coefficients $\beta_n$, also several spatially separated 
solitons can be described by the trial wave function \eqref{eq:ansatz}.
Evaluating the dynamical equations~\eqref{eq:TDVP-eq-of-motion} for the trial 
wave function~\eqref{eq:ansatz} yields after some calculation the  
equations of motion for the variational parameters
\begin{subequations}%
\begin{align}
	\label{eq:bgla}
	\dot{\alpha}_{n}&=-4\textup{i}\alpha_{n}^2+\textup{i}V_2^{n},\\
	\label{eq:bglb}
	\dot{\beta}_{n}&=-4\textup{i}\alpha_{n}\beta_{n}-\textup{i}V_1^{n},\\
	\label{eq:bglc}
\dot{\gamma}_{n}&=-2\textup{i}\alpha_{n}+\textup{i}\beta_{n}^2-\textup{i}V_0^
{n} \,,
\end{align}%
\label{eq:bgl}%
\end{subequations}
where $\edit{V_0^{n},V_1^{n}}$ \edit{and} $\edit{V_2^{n}}$ \edit{are solutions of the linear system of equations}
\begin{equation}
\label{eq:lgs}
 \begin{split}
 \edit{\sum\limits_{n=1}^{\Ng}\Braket{\frac{\partial\psi}{
 \partial\zz}|g_{n}\left(x^2V_2^{n}+xV_1^{n}+V_0^{n}\right)}}
%  &\\ 
 \edit{=-\left\langle\frac{\partial\psi}{\partial\zz}
 \bigg||\psi|^2\psi\right\rangle}&\,.
 \end{split}
 \end{equation} 
\strike{$\vec{v} = (V_0^{n},V_1^{n},V_2^{n})$ with
$n=1,2,\ldots,\Ng$ 
is the solution of the linear system of equations} \strikeeq{\cite{Rau2010a}}
\strikeeq{
\begin{equation}
\strike{ K \vec{v} = \vec{r}\,.}
\end{equation}
}
\strike{Here, }$\strike{K}$\strike{ and} $\strike{\vec{r}}$\strike{ are defined 
by}
\strikeeq{
\begin{subequations}
\begin{align}
\strike{
 K =
 \begin{pmatrix}
  \left<1  \right>_{kn} & \left<x  \right>_{kn} & \left<x^2\right>_{kn} \\
  \left<x  \right>_{kn} & \left<x^2\right>_{kn} & \left<x^3\right>_{kn} \\
  \left<-x^2\right>_{kn} & \left<-x^3\right>_{kn} & \left<-x^4\right>_{kn} 
 \end{pmatrix} ,}
 \\ \quad
\strike{
 \vec{r} = - \sum_{n=1}^{\Ng}
 \begin{pmatrix}
  \left<|   \psi|^2  \right>_{kn}\\
  \left<x|  \psi|^2  \right>_{kn}\\
  \left<-x^2|\psi|^2  \right>_{kn}
 \end{pmatrix} \,,
 }
\end{align}
\end{subequations}
}
% 
% \comment{Beide Gleichungen streichen}
\strike{where $\left<\cdot\right>_{kn}=\left<g_n|\cdot|g_k\right>$.}
% \comment{Diese Gleichung für Referee 2 genauer erklären.}\\
% \comment{Tobi: So wie ich das im Moment sehe ist die Formulierung von dem LGS wie wir es angegeben haben nicht richtig. Ich würde vorschlagen, wie ganz zu Beginn, die implizite From angeben, da man dabei ohne irgendwelche zusätzlichen Indizes einen leicht verständlichen Ausdruck erhält}
% 
Since the trial wave function $\psi$ is composed of Gaussian functions, all 
integrals occurring here are of the type $\int_{-\infty}^{\infty} x^d \,
\ue^{-\alpha x^2 + \beta x + \gamma} \, \ud x$ with $d=0,1,2,3,4$ and they are 
trivial to be calculated analytically.

In order to provide a better physical understanding, we note that the momentum 
$p_n$ and central position $x_{n}$ of each soliton are related to the 
variational parameters via 
\begin{subequations}
	\label{eq:variablen}
	\begin{align}
	\label{eq:p}
	p_n &= \text{Im}(\beta_{n})\,,\\
	\label{eq:x0}
	x_{n} &= \frac{\text{Re}(\beta_{n})}{2\alpha_{n}}\,.
	\end{align}
\end{subequations}

\subsubsection{Single-Gaussian approximation to the soliton}

%%%%%%%%%%%%%%%%%%%%%%%%%%%%%%%%%%%%%%%%%%%%%%%%%%%%%%%%%%%%%%%%%%%%%%%%%%%%%%%
\begin{figure}[t]
\includegraphics[width=\columnwidth]{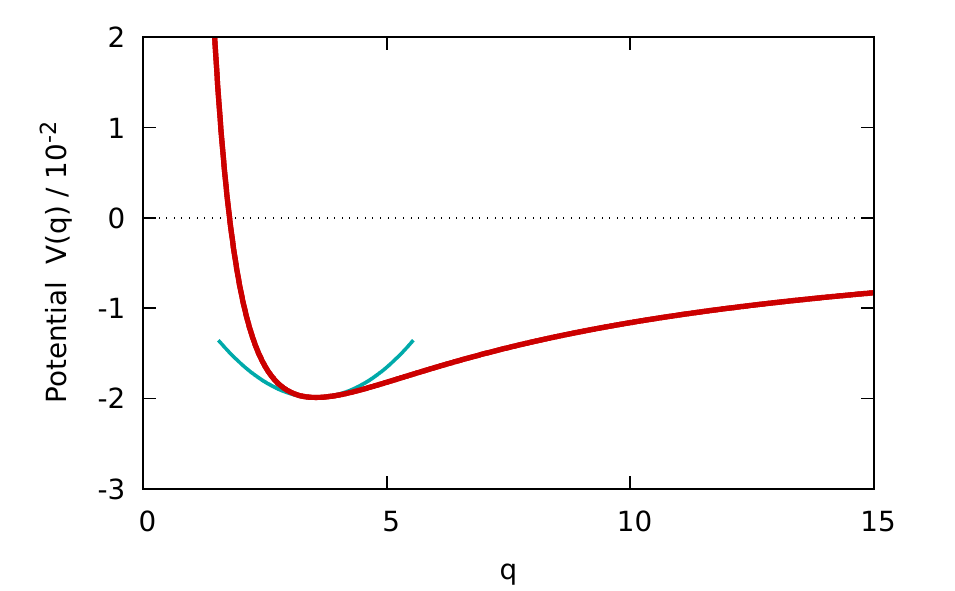}
\caption{%
Potential $V(q)$, \EQ~\eqref{eq:H1}, of the Hamiltonian picture 
describing a one-dimensional soliton in the framework of a single Gaussian 
variational approach (red line). 
The minimum represents the stable ground state of the soliton and the 
oscillation frequency for small excitations can be obtained from the harmonic 
approximation (blue line).
}
\label{fig:Ham-potential}
\end{figure}
%%%%%%%%%%%%%%%%%%%%%%%%%%%%%%%%%%%%%%%%%%%%%%%%%%%%%%%%%%%%%%%%%%%%%%%%%%%%%%%

From the equations of motion~\eqref{eq:bgl}, one cannot easily see that the 
Gaussian approach is able to reproduce the soliton ground state.
However, regarding the simplest approximation of a single soliton described by 
a single Gaussian wave function ($\Ng=1$), this is easily visible.
For this purpose, we define the generalized coordinate 
$q=\sqrt{\left \langle x^2 \right \rangle} = 1/(2\sqrt{\text{Re}\,\alpha_1})$ 
and its conjugate momentum 
$p={\text{Im}\,\alpha_1}/{\sqrt{\text{Re}\,\alpha_1}}$.
In this special case, the dynamical equations~\eqref{eq:bgl} and the energy 
functional \eqref{eq:TDVP-E} can be evaluated analytically and they are 
equivalent to Hamilton's equations corresponding to the Hamiltonian
\begin{equation}
	H(p,q)
	= T(p) + V(q)
	= p^2 \; + \; \frac{1}{4q^2} - \frac{1}{4 \sqrt{\pi} q} \,,
	\label{eq:H1}
\end{equation}
which defines a potential $V(q)$.
As shown in \FIG~\ref{fig:Ham-potential}, this potential exhibits a 
minimum (located at $q_{\textup{min}} = 2\sqrt{\pi}$) which corresponds to the 
soliton's ground state.
The form of the potential $V(q)$ also shows that the soliton is stable with 
respect to small excitations, and a harmonic approximation of the potential at 
its minimum yields the collective oscillation frequency of the atomic cloud.
In case of a description with a larger number of Gaussians, such a global 
Hamiltonian description is no longer possible, however this qualitative picture 
remains valid.

%%% Section: Results %%%%%%%%%%%%%%%%%%%%%%%%%%%%%%%%%%%%%%%%%%%%%%%%%%%%%%%
\section{Results}
\label{sec:results}

\subsection{Stationary solutions}
\label{subsec:stationary}

We begin by discussing the stationary ground state of a single soliton.
Due to the translational invariance of the system, it is appropriate to treat 
the system in its center of mass frame.
In this case, the soliton's center is located at $x=0$ and this also holds for 
all the coupled Gaussian wave functions, meaning that we can set $\beta_{n}=0$ 
throughout in this case. 
Stationary solutions within the variational approach are given if all the 
time-derivatives in \EQ~\eqref{eq:bgl} are zero [except the oscillating phase in 
\EQ~\eqref{eq:bglc}].
The resulting equations then form a nonlinear system of equations for the 
variational parameters which we solve analytically for a single Gaussian ($N_g=1$) and numerically otherwise after providing appropriate 
initial values.

%%%%%%%%%%%%%%%%%%%%%%%%%%%%%%%%%%%%%%%%%%%%%%%%%%%%%%%%%%%%%%%%%%%%%%%%%%%%%%
\begin{figure}
\includegraphics[width=\columnwidth]{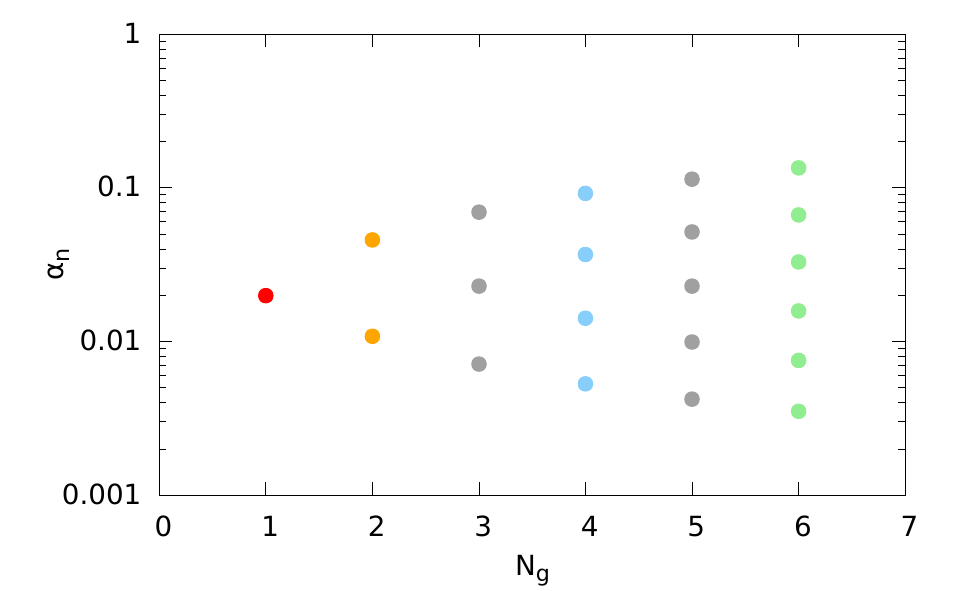}
\caption{%
Variational parameters $\alpha_{n}$ for the stationary ground state of a single 
soliton described by different numbers $\Ng$ of coupled Gaussians wave 
functions.
For the ground state, all the width parameters are real and result in a purely 
real wave function as expected from the analytical solution~\eqref{eq:PSI1}.
(The parameters $\gamma_{n}$ which determine the weight of each Gaussian are 
not shown here, and the colors of the dots for $\Ng=1,2,4,6$ correspond to the 
line colors in \FIG~\ref{fig:stat-states}.)
}
\label{fig:variationalparameter}
\end{figure}
%%%%%%%%%%%%%%%%%%%%%%%%%%%%%%%%%%%%%%%%%%%%%%%%%%%%%%%%%%%%%%%%%%%%%%%%%%%%%%

As the result of this root search, \FIG~\ref{fig:variationalparameter} shows 
the values of the variational width parameters $\alpha_{n}$ in 
dependence of the number $\Ng$ of coupled Gaussian wave functions.
The width parameters $\alpha_{n}$ of the trial wave function are all real in 
this case yielding a purely real ground state wave function as it is expected 
from the analytical solution~\eqref{eq:PSI1}.
In addition to the width parameters, the root search also provides the values of 
the parameters $\gamma_{n}$ which determine the weight of each Gaussian (not 
shown).
Using a single Gaussian $(\Ng=1)$, the value of 
$\alpha_{n}=1/(16\pi)$ 
is the 
best fit to the non-Gaussian soliton wave function \eqref{eq:PSI1}.
Adding a second, coupled Gaussian ($\Ng=2$), the tendency of the single 
contributions is that one of them corresponds to a larger width and the other to 
a smaller width. 
This trend continues with increasing number $\Ng$ of Gaussians, \ie~the 
smallest (largest) width parameter becomes smaller (larger) and the others 
cover the intermediate regime.

%%%%%%%%%%%%%%%%%%%%%%%%%%%%%%%%%%%%%%%%%%%%%%%%%%%%%%%%%%%%%%%%%%%%%%%%%%%%
\begin{figure}[t]
\includegraphics[width=\columnwidth]{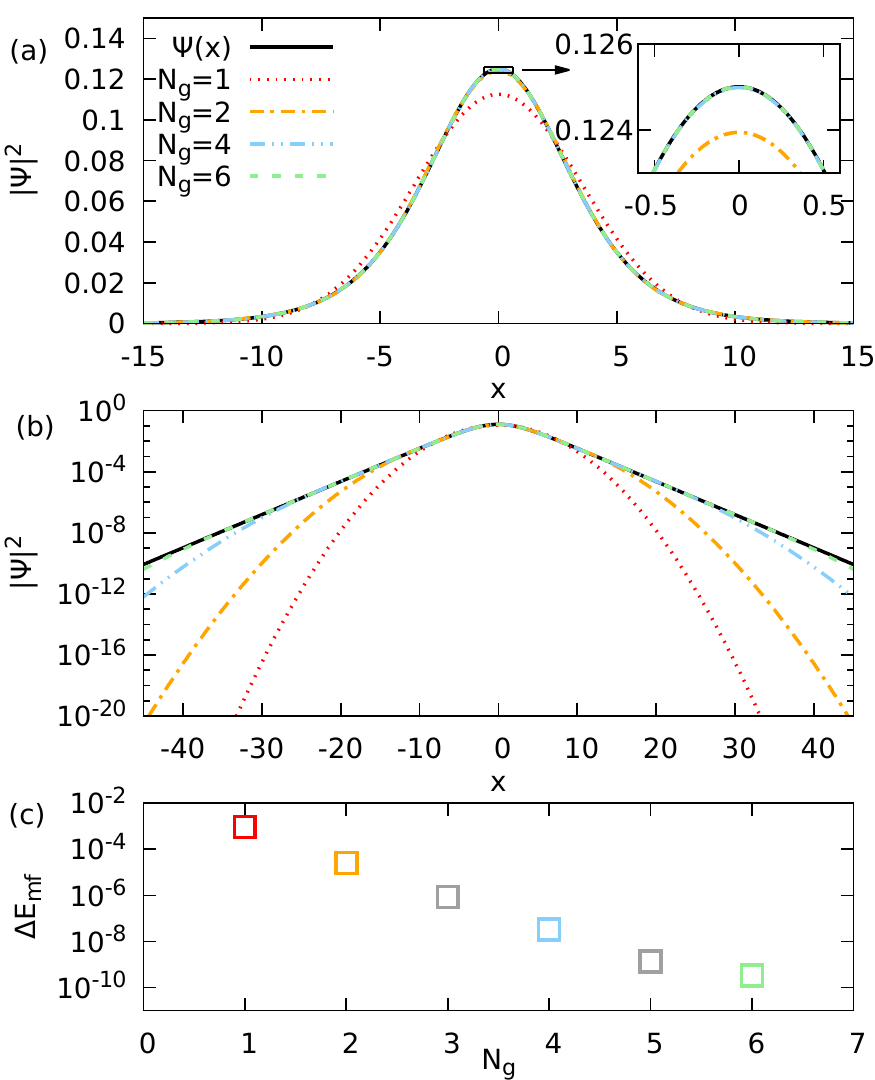} 
\caption{%
Comparison of the approximated wave functions within the variational approach 
(colored nonsolid lines) and the analytical solution~\eqref{eq:PSI1} (black 
solid line).
Panel (a) shows the wave function on a linear axis, illustrating the general 
behavior.
In panel (b) a logarithmic scale is used in order to demonstrate the heavy-tail 
approximation behavior.
Panel (c) shows the difference between the mean-field energy and the analytical 
ground state energy [$\Delta E_\text{mf} = E_\text{mf} + 1/48$ according to 
\EQ~\eqref{eq:E}].
An additional Gaussian in the trial wave function improves the ground state 
energy by roughly one to two orders of magnitude.
% 
% \comment{Tobi, kannst du bitte hier noch die Linienstile von ``$N_g=6$'' und 
% ``$\psi$'' austauschen, sodass $\psi$ als korrekte Lösung die durchgezogene 
% Linie bekommt. Evtl. dann die Reihenfolge ändern, sondass $\psi$ als erstes 
% geplottet wird.}
}
% \comment{Tobi: Erledigt}
\label{fig:stat-states}
\end{figure}
%%%%%%%%%%%%%%%%%%%%%%%%%%%%%%%%%%%%%%%%%%%%%%%%%%%%%%%%%%%%%%%%%%%%%%%%%%%%

In \FIG~\ref{fig:stat-states}, we show the soliton wave functions which 
correspond to the width parameters from \FIG~\ref{fig:variationalparameter}.
The colored solid lines present the results of the variational approach and the 
dashed black line is the analytic result~\eqref{eq:PSI1}.
As it becomes obvious in \FIG~\ref{fig:stat-states}(a), a single Gaussian 
($\Ng=1$; red line) qualitatively reproduces the correct soliton wave 
function, while it quantitatively fails to reproduce the higher density at the 
center of the soliton.
A number of $\Ng=2$ Gaussians (orange line) already improves the wave 
function significantly over the whole spacial range; the density peak compared 
with the pure Gaussian form is reproduced better and also the tendency of a 
more 
rapid decrease of the density away from the center is visible.
For even higher number of Gaussians ($\Ng=4$; blue line and $\Ng=6$; 
green  line), the true wave function is approximated better and better and 
a difference to the variational approach is no longer visible within the 
linewidth of the plot [see the inset in \FIG~\ref{fig:stat-states}(a)].

As we demonstrate in \FIG~\ref{fig:stat-states}(b), this behavior not only 
holds at the center of the soliton but also for its tail:
The pure Gaussian variational approach ($\Ng=1$) clearly fails to reproduce the 
heavy-tail density distribution of the soliton. 
This is clearly expected since the Gaussian trial wave function decays like 
$\exp(-x^2)$ while the true soliton \eqref{eq:PSI1} only decays like $\exp(-x)$ 
for $x\to\infty$.
Far away from the soliton's center, the difference between the true and the 
trial wave function is, therefore, several orders of magnitude.
Increasing the number of Gaussian, the tail-approximation soon becomes better 
with improvements of several orders of magnitude for each step of $\Ng$.
Finally, only $\Ng=6$ coupled Gaussians are sufficient to reproduce the correct 
tail behavior over a very broad range.

The improvement of the wave function's approximation with increasing number of 
coupled Gaussians is also reflected in the value of the mean-field energy which 
one obtains from the variational approach.
To show this, we present in \FIG~\ref{fig:stat-states}(c) the difference 
between the ground state energy within the variational approach and the value 
\eqref{eq:E} of the analytical solution, $\Delta E_\text{mf} = E_\text{mf} + 
1/48$.
For a single Gaussian, this difference is $\Delta E_\text{mf} = 
9.39\times10^{-4}$, but it decreases by about one to two orders of magnitude 
with each additional Gaussian.
For $\Ng=6$ coupled Gaussian, the ground state energy is approximated very well 
with an error only of the order of $10^{-10}$.
This, again, proves that the variational approach even with a low number of 
Gaussians is highly appropriate in order to approximate the soliton wave 
function.

\subsection{Dynamics and collisions}

As we have seen in \SEC~\ref{subsec:stationary} the variational approach 
converges rapidly to the analytical stationary bright soliton solution for 
increasing number of Gaussians. 
In this section, we address the question to what extent the variational 
approach is capable of describing the dynamics of the system. 
Therefore, we investigate the dynamics of collisions between several solitons in 
the following.

\subsubsection{Collisions of two solitons} 

As a simple case, we first consider two colliding solitons.
For this case, we use one and two Gaussians per soliton, and describe the 
collision in its center-of-mass frame.
Under the assumption that the colliding solitons consist of the same number of 
particles, they both have the same momentum and the same distance to the origin 
of the center-of-mass frame,
\begin{subequations}
\begin{gather}
x_{1}=-x_{2}\equiv x_0\,,\\
p_2=-p_1 \equiv p\,.
\end{gather}
\end{subequations}
For the initial soliton wave function, we take in the variational approach 
\EQ~\eqref{eq:ansatz} with an appropriate set of variational parameters and for 
the grid calculations, we discretize the wave function of 
each individual soliton according to
\begin{equation}
  \psi(x, t=0)=\sqrt{\frac{1}{8}} \, 
  \frac{\ue^{\textup{i}p(x-x_0)+\textup{i}\varphi}}{\cosh[(x-x_0)/4]}\,,
  \label{eq:PSI3}
\end{equation}
where $x_0$ is the central position of the soliton, and $p, \varphi$ are the 
initial momentum and phase, respectively.
Configurations of several solitons can also be initialized by 
\EQ~\eqref{eq:PSI3} if their overlap is negligible. 
The global phase of the solitons is a free parameter and not measurable.
Only their relative phase matters which we take into account by setting the 
phase 
of the first soliton to zero, while introducing a phase shift $\Delta\varphi$ 
between the two solitons,
\begin{subequations}
	\begin{align}
	\varphi_1&=0\,,\\
	\varphi_2&=\Delta \varphi\,.
	\end{align}
\end{subequations}
By an appropriate choice of $\Delta\varphi$ the wave functions can be set, 
inter alia, symmetric ($\Delta\varphi=0$) or antisymmetric 
($\Delta\varphi=\pi$). 
Note that the initial separation of the solitons needs to be chosen large 
enough so that the overlap of the solitons is negligible at the beginning. 
With these choices, the only remaining physical parameters in the soliton 
collisions are their phase difference $\Delta\varphi$ and their initial momentum 
$p$.
In the numerical simulations, the number of Gaussians $\Ng$ is a further 
computational input parameter.

%%%%%%%%%%%%%%%%%%%%%%%%%%%%%%%%%%%%%%%%%%%%%%%%%%%%%%%%%%%%%%%%%%%%%%%%%%%%%%%%
\begin{figure*}[t]
\includegraphics[width=\textwidth]{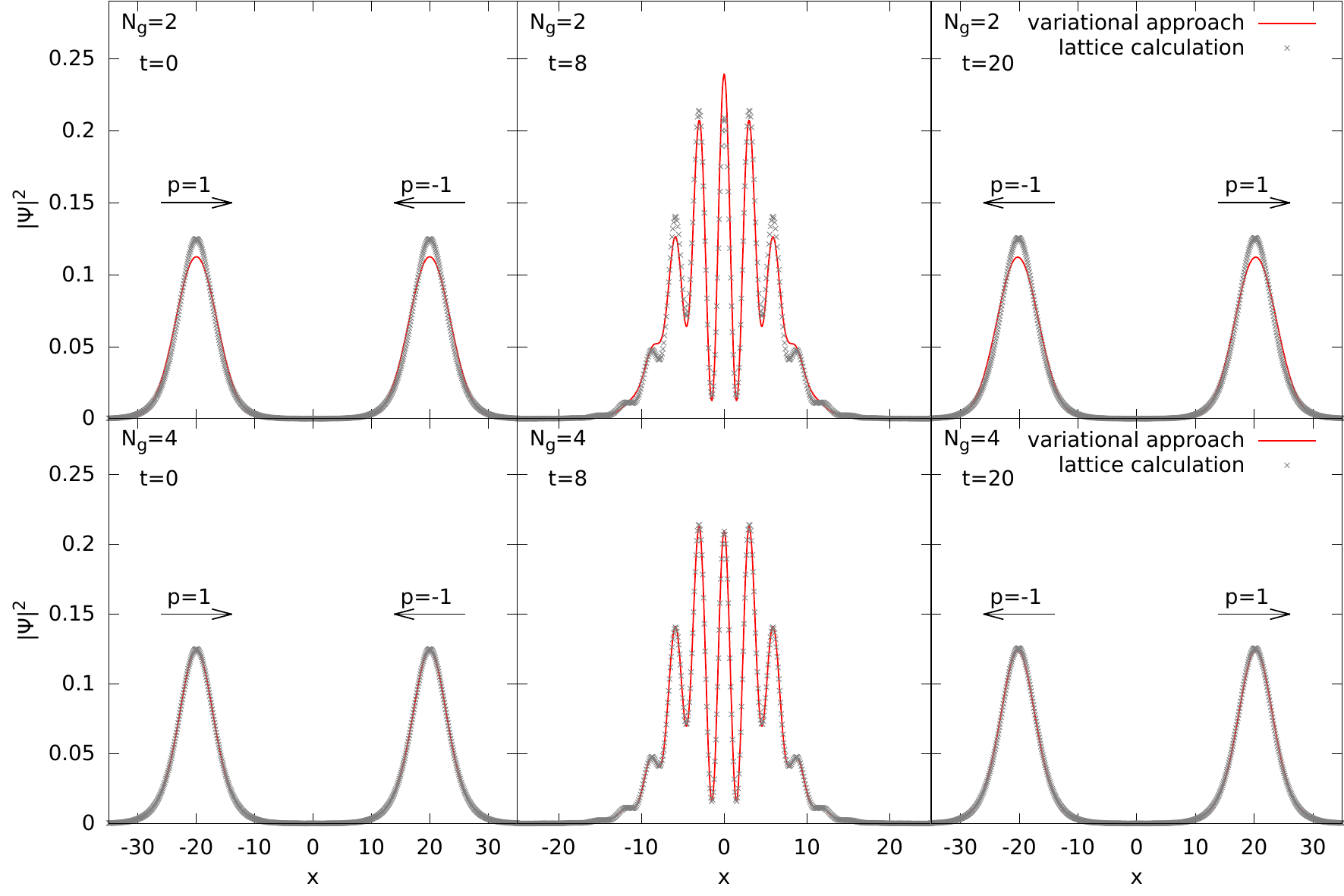}
\caption{%
Collision of two solitons with initial momentum $|p|=1$ and phase difference 
$\Delta\varphi=0$.
The comparison is shown between 
the results obtained with the variational approach (red lines) 
with $\Ng=2$ (top row; one Gaussian per soliton) and $\Ng=4$ (bottom row; two 
Gaussians per soliton) and the numerically exact dynamics (gray dots).
The wave functions are plotted before ($t=0$), during ($t=8$), and after 
($t=20$) the collision. 
}
\label{fig:comparison1}
\end{figure*}
%%%%%%%%%%%%%%%%%%%%%%%%%%%%%%%%%%%%%%%%%%%%%%%%%%%%%%%%%%%%%%%%%%%%%%%%%%%%%%%%

\paragraph{\edit{High-energy collisions}}

In \FIG~\ref{fig:comparison1}, we consider the \edit{high-energy} collision of 
two fast moving 
solitons, where we define 'fast' with respect to the kinetic energy of the 
soliton in its center-of-mass frame, where it is 
$ \langle p^2\rangle= 
\left\langle \psi |-\partial_x^2  | \psi \right\rangle={1}/
{48}$
with $\psi$ in \EQ~\eqref{eq:PSI1}.
From this, we take 
\begin{equation}
  \prest = \sqrt{\frac{1}{48}}\approx 0.144
  \label{eq:p-value}
\end{equation}
as a reference for the magnitude of the soliton momentum and to classify a 
high- 
and low-energy regime of the dynamics.
We compare the results obtained from the 
variational approach (red lines) and from numerically exact grid calculations 
(gray dots). 
For the initial momentum we set $p=1$ which is significantly larger than  
$\prest$ in \EQ~\eqref{eq:p-value}. 
The phase difference between the solitons is set to $\Delta\varphi=0$ which 
implies that the wave function is symmetric.

In the top row of \FIG~\ref{fig:comparison1} only a single Gaussian 
per soliton is used (red line; $\Ng=2$ altogether).
At $t=0$ both solitons are well separated in space so that their overlap can be 
neglected and they approach each other with the same momentum $p=1$, 
respectively.
As we have already discussed in the stationary case, the simple 
variational approach qualitatively reproduces the exact soliton solution, but 
small quantitative differences are present.
At time $t=8$ the solitons strongly overlap and interfere, resulting in several 
density peaks. 
It is clearly visible that both the variational approach as well as the 
numerically exact solution show the same behavior. 
After the collision ($t=20$; third column) the solitons move away from each 
other in the same state of motion as before the collision. 
\edit{
Between the variational approach and the application of an approximative 
time-evolution operator, no differences in the shape and height of the solitons 
before and after the collision are visible, which shows that the latter is an 
appropriate method to integrate the dynamics.
}

We note that the computational effort to obtain these results shows significant 
differences between both approa\-ches: 
The variational approach is about 75 times faster than the lattice calculation 
while still maintaining nearly the same quantitative behavior.

In the bottom row of \FIG~\ref{fig:comparison1}, the same situation is shown 
using $\Ng=4$ coupled Gaussians (two Gaussians per soliton).
The improvement in the variational approach is evident.
The collision 
dynamics in the variational framework and the numerically exact solution can no 
longer be distinguished although the variational approach is still five times 
faster than the lattice calculation.
Even at the time of high spacial overlap, the variational approach perfectly 
reproduces the soliton wave functions.
It is furthermore remarkable that, although there are $\Ng=4$ 
Gaussian wave functions involved in the simulation process, they form 
two well-separated solitons also after their collision.

By increasing the initial momentum of the colliding solitons (not 
shown) the amount of peaks during the overlapping regime increases.
Changing the phase difference between the solitons to a value different than 
$\Delta\varphi=0$ or $\Delta\varphi=\pi$ also the symmetry in the collision 
dynamics gets lost. However, the variational approach remains capable of 
describing the behavior correctly. 
\strike{We note that this changes if the momentum is comparable or lower than 
the rest momentum of the solitons, }$\strike{p\lesssim\prest}$. 
\strike{Here, the results of the variational approach can significantly differ from 
the grid calculations. 
This is a common phenomenon in coupled Gaussian approaches if the single wave 
functions become too close in configuration and momentum space.
This can be overcome \eg~by applying constraints to the dynamical 
equations }\cite{Fabcic2008a}.

\paragraph{\edit{Low-energy collisions}}

\begin{figure*}[t]
\includegraphics[width=\textwidth]{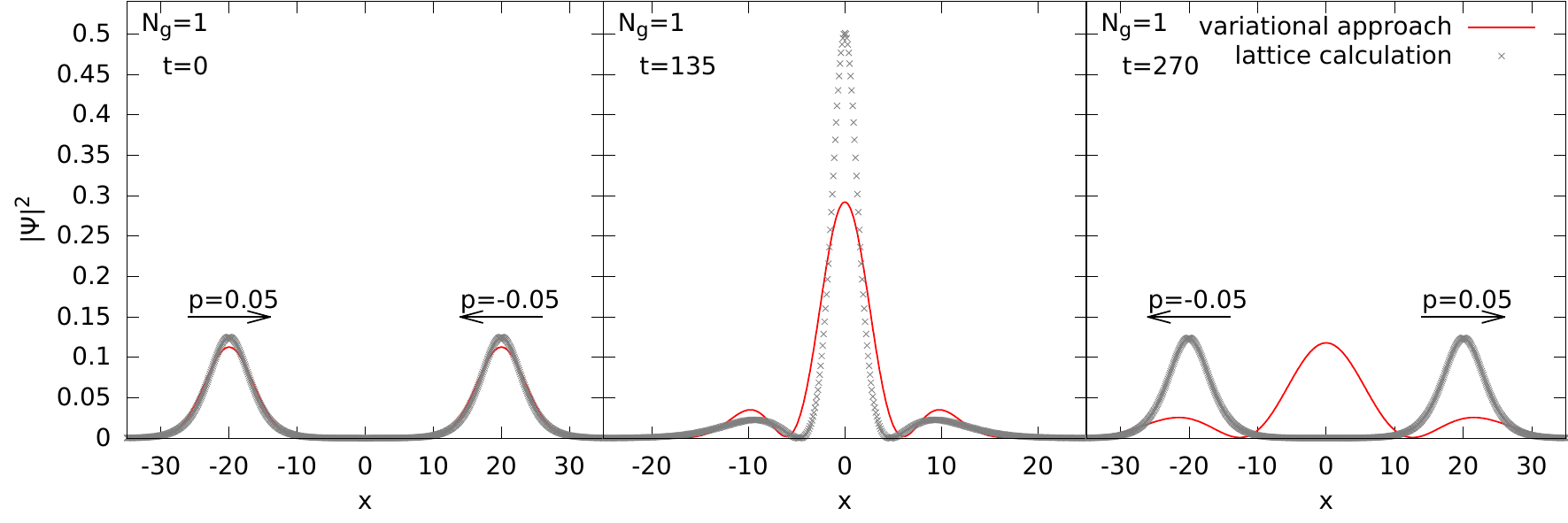}
\caption{%
\edit{Collision of two solitons with initial momentum $|p|=0.05$ 
and phase difference $\Delta\varphi=0$. The dynamics for the 
variational approach (red line) was calculated for $\Ng=2$ 
Gaussians (one Gaussian per soliton) and is compared with the results of the 
lattice calculations (gray dots).}
}
\label{fig:lowmomentum}
\end{figure*}

\edit{
We have seen that the variational approach is capable of describing the 
dynamics of 
the solitons correctly in the high-energy regime. 
Next we consider collisions where the momentum of the solitons is lower than 
their rest momentum $\edit{\prest}$.
The collision of two solitons with a momentum of $p=0.05 \ll \prest$ 
is shown in \FIG~\ref{fig:lowmomentum}. 
% The latter is significantly smaller than the rest momentum of $p=0.144$. 
The 
phase difference between the solitons is set to $\Delta\varphi=0$ 
and the variational approach uses $\Ng=2$ Gaussians (one 
Gaussian per soliton). 
}

\edit{
Since we are plotting the modulus squared of the wave function, there is no 
difference visible at $t=0$ between \FIGS~\ref{fig:comparison1} and 
\ref{fig:lowmomentum}. 
At $t=135$, during the collision, the solitons overlap strongly and the wave 
function's amplitude within the variational approach is significantly smaller 
than expected from the grid calculations.
However, the general shape of the interfering solitons, \ie~the main peak at 
$x=0$ and two side maxima at $x\approx \pm10$, is clearly reproduced by the 
variational approach. 
Significant differences between the two descriptions become obvious at 
$t=270$. 
Here, the grid calculation shows two well-separated, departing solitons. The 
solitons show the same behavior as they did during collusions with high 
momentum. 
By contrast, within the variational approach the wave packet has not 
yet separated into different solitons (red line). Instead, they have partially 
merged.
%and are no longer seperated after the collision.   
It takes the solitons described by the variational approach another time span of 
$\Delta t=230$ to separate. 
After they have separated the solitons oscillate and their shape is not 
invariant any more (not shown in the figure). 
Note that during the whole simulation, the total energy as well as the norm 
of the wave function is conserved.
}
%\edit{
 %There are also scenarios, in that the conservation of the total energy and %the norm of the wave function is violated. By adding a phase difference %between the solitons other than }$\edit{\Delta\varphi=0}$\edit{ or %}$\edit{\Delta\varphi=\pi}$\edit{ the variational approach is not able to %provide a physically meaningful result. The solitons collide but are not able %to separate. After a short time the wave function dissolves while the total %energy and the norm are not conserved any more.}

\edit{
We have carefully checked that the observed merging effect also occurs when 
increassing the number of coupled Gaussians, so that it is not an artifact of 
an insufficient choice of the trial wave function.
%  One might expect that this behavior is due to the restriction of the 
% variational space to a single Gaussian wave function per soliton.
% By contrast, the observation also occurs if the number of trial wave 
% functions is increased.
% However, our calculations have shown that adding more Gaussians does not 
% improve the results. 
However, from the fact that the merge only occurs in the low-energy regime, we 
conclude  that it is rather a numerical effect due to
% that the parameter regime 
% in which the merging of solitons occurs is enhanced in the variational 
% description, and we expect this shift to result from 
the closeness of the parameters in variational space.
It is a well-known phenomenon in coupled Gaussian approaches that singularities 
in the dynamical equations \eqref{eq:TDVP-eq-of-motion} occur if two (nearly) 
identical 
Gaussians contribute to the total wave function.
% As a consequence, dynamical effects can be enhanced in case that variational 
% parameters are sufficiently close to each other.
This is clearly the case for low-energy collisions where both position and 
velocity of the soliton are almost identical during the collision.
% the single wave functions become too close in configuration and momentum 
% space.
}

%%%%%%%%%%%%%%%%%%%%%%%%%%%%%%%%%%%%%%%%%%%%%%%%%%%%%%%%%%%%%%%%%%%%%%%%%%%%%%%
\begin{figure*}[t]
\includegraphics[width=\textwidth]{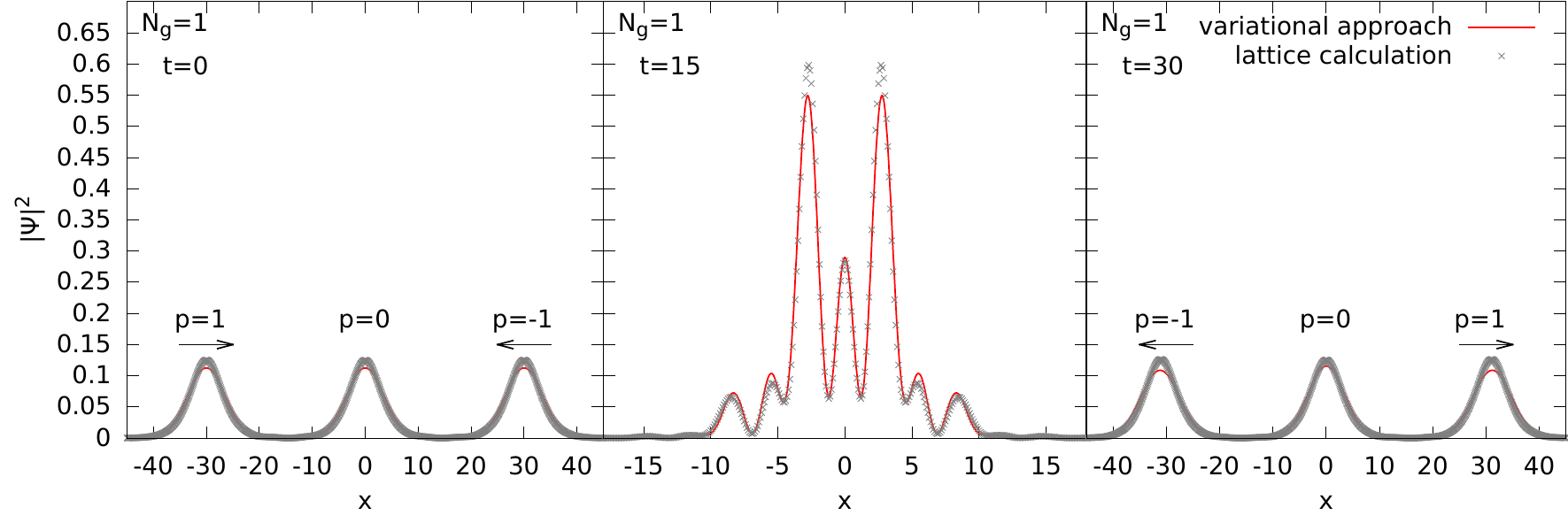}
\caption{%
Collision of three solitons in the center-of-mass frame. 
Shown is the modulus squared of the wave function obtained from the 
variational approach with $\Ng=6$ (red lines; two Gaussians per soliton) and 
the numerically exact dynamics (gray dots). 
The two moving solitons have the same momentum $p=\pm1.5$ and there is no phase 
difference between them ($\Delta \varphi=0$). 
}
\label{3solitons}
\end{figure*}
%%%%%%%%%%%%%%%%%%%%%%%%%%%%%%%%%%%%%%%%%%%%%%%%%%%%%%%%%%%%%%%%%%%%%%%%%%%%%%%

\subsubsection{Collision between several solitons}

We have seen, that the variational approach produces excellent results 
for the collision of two solitons at high momentum. 
Next we consider the dynamics of three solitons where two moving solitons 
approach a soliton at rest.
The moving solitons have the same momentum and accordingly hit the third 
soliton 
at the same time. 
We place the resting soliton in the center of coordinates so that we are again 
in the center-of-mass frame. 
Furthermore, the phase difference between the moving solitons is set to zero, 
so that the wave function is symmetric. 
In \FIG~\ref{3solitons} the collision of three solitons is shown with the 
moving 
solitons initially having a momentum of $p=\pm1.5$. The variational approach for 
$\Ng=6$ (two Gaussians per soliton) is compared with the results of lattice 
calculations.

Before the collision (left frame) barely any difference between the variational 
approach and the lattice calculation is visible. 
This changes slightly during the collision (center frame). 
While the variational approach reproduces the shape of the lattice 
calculations very well in the surroundings of the two maxima, the maxima 
themselves 
slightly differ from the lattice calculations.
The two moving solitons penetrate the one at rest and all three solitons are in 
the same state of movement as before (right frame). 
The small differences during the collision do not affect the behavior after 
it. 
As in the first frame, no differences between the variational approach and the 
lattice calculations are visible on the linear scale. This shows that the 
variational approach is capable of producing excellent agreement even for 
several 
soliton interactions.

Our simulations show that the Gaussian variational approach 
correctly describes collisions of even more solitons (not presented in this 
paper).
For investigations of the corresponding dynamics for odd wave functions as well 
as animations, we refer the reader to \REF~\cite{ilg}.

%%% Section: Conclusions %%%%%%%%%%%%%%%%%%%%%%%%%%%%%%%%%%%%%%%%%%%%%%%%%%%
\section{Conclusion}

In this paper, we have investigated solitons in the one-dimensional GPE within 
an extended variational approach using coupled Gaussian wave functions.
We have demonstrated that the variational approach perfectly 
reproduces the ground state of the soliton concerning the shape of the wave 
function as well as the ground state energy.
This observation is especially remarkable because the soliton wave function 
differs from the pure Gaussian form with regard to its heavy-tail decay and the 
increased density at its center.
The time-dependent variational approach applied in this paper also well 
describes the soliton dynamics in the high-energy regime.
Already a small number of trial wave functions are sufficient to reproduce the 
time-dependent density distribution to a very high accuracy.
\edit{%
By contrast, in the low-energy regime differences between the numerical and 
the variational approach can be observed which we expect to result from 
numerical issues due to the almost identical contribution of 
different wave functions during the low-energy collisions.
}

\edit{
We note that the methods described in this paper are directly applicable to 
higher-dimensional systems with more complicated interactions \cite{Eichler1, 
Eichler2} and that our results agree with previous studies in similar cases 
using perturbation theory \cite{Michinel1999}.
Our future work will also take into account the low-energy regime of 
slowly moving solitons in more detail involving Gaussian constraints in the 
dynamical equations \cite{Fabcic2008a}.
}

% \section*{Author contribution statement}

%%% Bibliography %%%%%%%%%%%%%%%%%%%%%%%%%%%%%%%%%%%%%%%%%%%%%%%%%%%%%%%%%%%%
% \newpage
% \bibliographystyle{epj}
% \bibliography{literature.bib}

\end{document}